\documentclass[amsmath,amssymb,aps,prl,reprint,floatfix,superscriptaddress]{revtex4-2}

\usepackage{graphicx}
\usepackage{dcolumn}
\usepackage{bm}
\usepackage[version=4]{mhchem}
\usepackage[breaklinks,colorlinks,linkcolor={blue}, %
            citecolor={blue},urlcolor={blue}]{hyperref}

\begin{document}

\title{Nonunique fraction of Fock exchange for defects in two-dimensional materials}
\author{Wei Chen} 
\affiliation{Institute of Condensed Matter and Nanoscicence (IMCN), Universit\'e catholique de Louvain, Louvain-la-Neuve 1348, Belgium}
\author{Sin\'ead M.\ Griffin}
\affiliation{Molecular Foundry Division, Lawrence Berkeley National Laboratory, Berkeley, California 94720, USA}
\affiliation{Materials Sciences Division, Lawrence Berkeley National Laboratory, Berkeley, California 94720, USA}
\author{Gian-Marco Rignanese} 
\affiliation{Institute of Condensed Matter and Nanoscicence (IMCN), Universit\'e catholique de Louvain, Louvain-la-Neuve 1348, Belgium}
\author{Geoffroy Hautier} 
\affiliation{Institute of Condensed Matter and Nanoscicence (IMCN), Universit\'e catholique de Louvain, Louvain-la-Neuve 1348, Belgium}
\affiliation{Thayer School of Engineering, Dartmouth College, Hanover, New Hampshire 03755, USA}
\date{\today}

\begin{abstract}
By investigating the vacancy and substitutional defects in monolayer \ce{WS2} with hybrid functionals,
we find that there is no unique amount of Fock exchange that concurrently
satisfies the generalized Koopmans' condition and reproduces the band gap and band-edge positions.
Fixing the mixing parameter of Fock exchange based on the band gap can lead to qualitatively 
incorrect defect physics in two-dimensional materials. 
Instead, excellent agreement is achieved with both experiment and many-body perturbation theory
within the $GW$ approximation once the mixing parameters are tuned individually for the defect species and the band edges.
We show the departure from a unique optimized mixing parameter is inherent to two-dimensional systems 
as the band edges experience a reduced screening whilst the localized defects are subject to
bulklike screening.
\end{abstract}

\maketitle

Two-dimensional (2D) materials are rapidly gaining ground in a range of technologies
owing to both their exotic electronic and optical properties due to quantum confinement~\cite{Duan2015}
and the increasing ease of manipulation of their properties on a layer-by-layer basis. 
Defects have been shown to be viable active sites for enabling electrocatalysis (e.g.,
hydrogen evolution~\cite{Ye2016,Ouyang2016,Tsai2017}) on otherwise inert basal planes of 2D materials.
In quantum information science, 
2D materials hold exceptional promise as defect hosts because their associated point defects 
typically present high spin states and deep defect levels~\cite{Tran2016,Caldwell2019,Mendelson2021}, 
and any near-surface defects are easier to manipulate and characterize.
The nature of such quantum defects is often elusive from experiment,
and reliable theoretical modeling is required to understand their atomistic origins~\cite{Turiansky2019,Ping2021}.
First-principles defect calculations are now typically carried out with hybrid density functionals,
the accuracy of which is closely related to the amount of Fock exchange
admixed with semilocal exchange in the framework of generalized Kohn-Sham density-functional theory (DFT)~\cite{Freysoldt2014}.
The mixing parameter of hybrid functionals is commonly determined on the basis of
the band gap of host systems from experiment or higher levels of theory.
It has been well validated that such band-gap optimized mixing parameters lead to
accurate defect levels for localized point defects in bulk materials~\cite{Chen2013}.

The success of hybrid functionals can be attributed to the fulfillment of generalized Koopmans' condition,
which for a localized defect, requires that the ionization energy and electron affinity to be equal, hence
recovering the exact piecewise linearity of the total energy upon electron occupation~\cite{Perdew1982,Lany2009,Stein2012,Perdew2017}.
Indeed, one can often find a single mixing parameter that describes the host band gap
and the defect localization equally well in bulk materials~\cite{Miceli2018}.
The uniqueness of optimal mixing parameter promotes the use of a fixed, band-gap targeted mixing parameter
in state-of-the-art defect calculations~\cite{Freysoldt2014}.
The same approach has been routinely applied to understand the defect properties in 2D materials~\cite{Li2016,Smart2018,Li2022,Li2022a}.
In this Letter, we show that, in contrast to bulk 3D materials,
there exists no unique amount of Fock exchange that can reproduce the experimental band gap and
satisfying Koopmans' conditions for the defect level concurrently.
Admixing a single fraction of Fock exchange thus could entail large errors in defect energy levels for 2D materials.
We show that imposing appropriate mixing parameters individually for defects and host band edges
is needed for an accurate description of defect energy levels,
and discuss the failure of the typical hybrid-functional defect computation scheme.

To illustrate this issue, we investigate various point defects in monolayer (ML) \ce{WS2},
including the sulfur monovacancy ($V_\text{S}$),
the single cobalt atom substituting sulfur (Co$_\text{S}$) or tungsten (Co$_\text{W}$),
and the single carbon atom substituting sulfur (C$_\text{S}$).
Several of these defects have recently become the subject of increasing theoretical and experimental efforts 
following the demonstrated quantum light emitters in
2D transition-metal dichalcogenides (TMDs)~\cite{He2015,Srivastava2015,Koperski2015,Chakraborty2015,PalaciosBerraquero2017,Schuler2020}, 
and so are well-motivated for further consideration here~\cite{Schuler2019,Zhang2019,Cochrane2021,Li2022}.

\begin{figure*}
\includegraphics{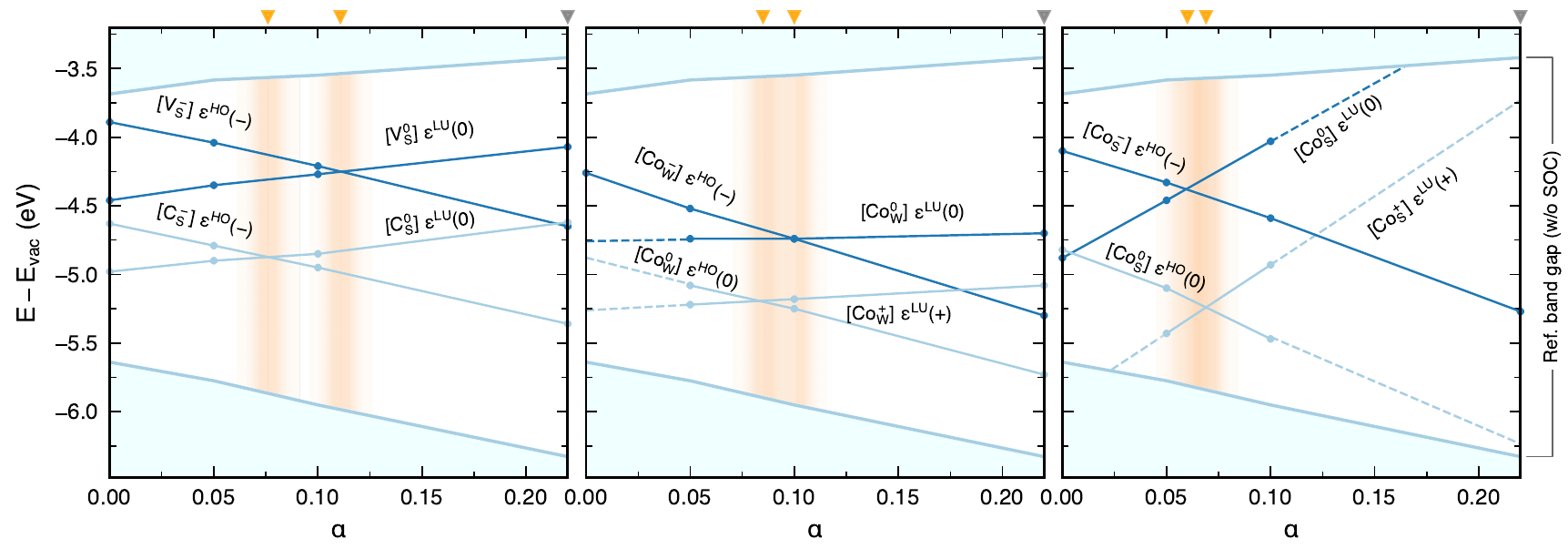}
\caption{\label{fig:be_ks}Single-particle defect levels as a function of mixing parameter $\alpha$
for various defects in the ML \ce{WS2}.
The eigenvalues are evaluated at the four representative $\alpha$ values shown by the solid dots and interpolated in between (solid lines).
Linear extrapolations (indicated by dashed lines) are used when the localized defect state cannot be stabilized within the band gap.
The band edges of ML \ce{WS2} are indicated by the blue shades.
All energies are referred to the vacuum level.
The orange marker indicates the mixing parameter $\alpha_\text{K}$ at which the generalized Koopmans’ condition
$\varepsilon^\text{HO}$($q$)$=\varepsilon^\text{LU}$($q+1$) is fulfilled,
whereas the grey marker indicates the optimal mixing parameter ($\alpha_\text{G}=0.22$) for reproducing
the band gap of ML \ce{WS2}.
Spin-orbit coupling (SOC) is not taken into account.}
\end{figure*}

Our analysis of Koopmans' condition is primarily carried out using a global hybrid functional
in which a single mixing parameter ($\alpha$) controls the amount of Fock exchange 
admixed with the semilocal Perdew-Burke-Ernzerhof (PBE) exchange~\cite{Perdew1996}.
At $\alpha=0.25$ this global hybrid functional is essentially the PBE0 functional~\cite{Perdew1996a,Adamo1999}.
Conventionally we refer to this family of one-parameter hybrid functionals as PBE0($\alpha$).
All hybrid-functional calculations are performed with the projector-augmented-wave method~\cite{Bloechl1994} 
as implemented in \textsc{vasp}~\cite{Kresse1996a,Kresse1996}.
As a higher level of theory, many-body perturbation theory within the $GW$ approximation~\cite{Hedin1965,Hybertsen1986} 
is used to establish an accurate reference for the hybrid functional.
Specifically, on top of a PBE starting point, one-shot $G_0W_0$ calculations are performed 
for the ML \ce{WS2} and some selected defects.
The truncated Coulomb interaction~\cite{IsmailBeigi2006} and the subsampling technique~\cite{Jornada2017} 
are used to address the slow convergence of quasiparticle (QP) energies specific to 2D systems~\cite{Qiu2016}.
The $G_0W_0$ calculations are performed with \textsc{berkeleygw}~\cite{Deslippe2012} 
interfaced to the DFT code from Quantum-\textsc{espresso}~\cite{Giannozzi2017}.
The details of computational parameters for the hybrid-functional and $G_0W_0$ calculations 
are provided in the Supplemental Material (SM) ~\footnote{
See Supplemental Material for computational details, finite-size corrections, and
Koopmans' condition with monolayer boron nitride, 
which includes Refs.~\cite{Schutte1987,Lin2016,Hamann2013,vanSetten2018,Godby1989}}.

We first survey the electronic structure of the pristine ML \ce{WS2}.
Obtained from exfoliating 2\textit{H}-\ce{WS2}, the ML \ce{WS2} is a direct-gap semiconductor
with a $K$--$K$ transition.
As a result of the trigonal prismatic crystal field splitting, 
the valence band maximum (VBM) is of $d_{x^2-y^2}$ and $d_{xy}$ character
whereas the conduction band minimum (CBM) is of $d_{z^2}$ character.
The $G_0W_0$ band gap of the ML \ce{WS2} is 2.9~eV without SOC.
To reproduce this reference $G_0W_0$ band gap, we find that the mixing parameter 
(dubbed as $\alpha_\text{G}$) needs to be adjusted to 0.22 for the hybrid functional.
{Once the SOC is included, this PBE0($\alpha_\text{G}$) functional leads to a band gap of 2.58~eV,
in good agreement with the experimentally reported values ranging from 2.4 to 2.7~eV~\cite{Chernikov2014,Schuler2019,Zhu2015}.
We note that the experimental determination of the band gap is subject to the 
dielectric screening effect arising from the substrate~\cite{Naik2018,Zibouche2021}.
}

To identify the mixing parameter fulfilling the generalized Koopmans' condition 
for a given localized defect $D$,
we calculate the single-particle eigenvalue $\varepsilon^\text{HO}(q)$ associated with the highest occupied (HO) state 
at charge state $q$ as well as the eigenvalue $\varepsilon^\text{LU}(q+1)$ of the lowest unoccupied (LU) state at charge state $q+1$.
The optimal mixing parameter $\alpha_\text{K}$ is determined when the equality 
$\varepsilon^\text{HO}(q)=\varepsilon^\text{LU}(q+1)$ holds.
To calculate this we consider  candidate defects  embedded in an orthorhombic supercell of 90 atoms
representing the ML \ce{WS2}.
In particular, we refer the band edges and single-particle defect eigenvalues to the vacuum level,
which is a natural and physical choice for 2D systems~\cite{Smart2018}.
While the alignment is straightforward for neutral defects,
the single-particle levels of charged defects are subject to finite-size effect arising from the spurious 
electrostatic interactions~\cite{Freysoldt2009,Komsa2012,Chen2013,Komsa2013}.
The alignment is further complicated by the fact that the electrostatic potential in the vacuum region
experiences an artificial bending due to the dipole moment introduced by the neutralizing charge background.
Here we apply the potential correction scheme of Chagas da Silva \textit{et al.}~\cite{ChagasdaSilva2021},
thereby achieving well-defined single-particle eigenvalues with respect to vacuum for charged defects.

\begin{figure*}
\includegraphics{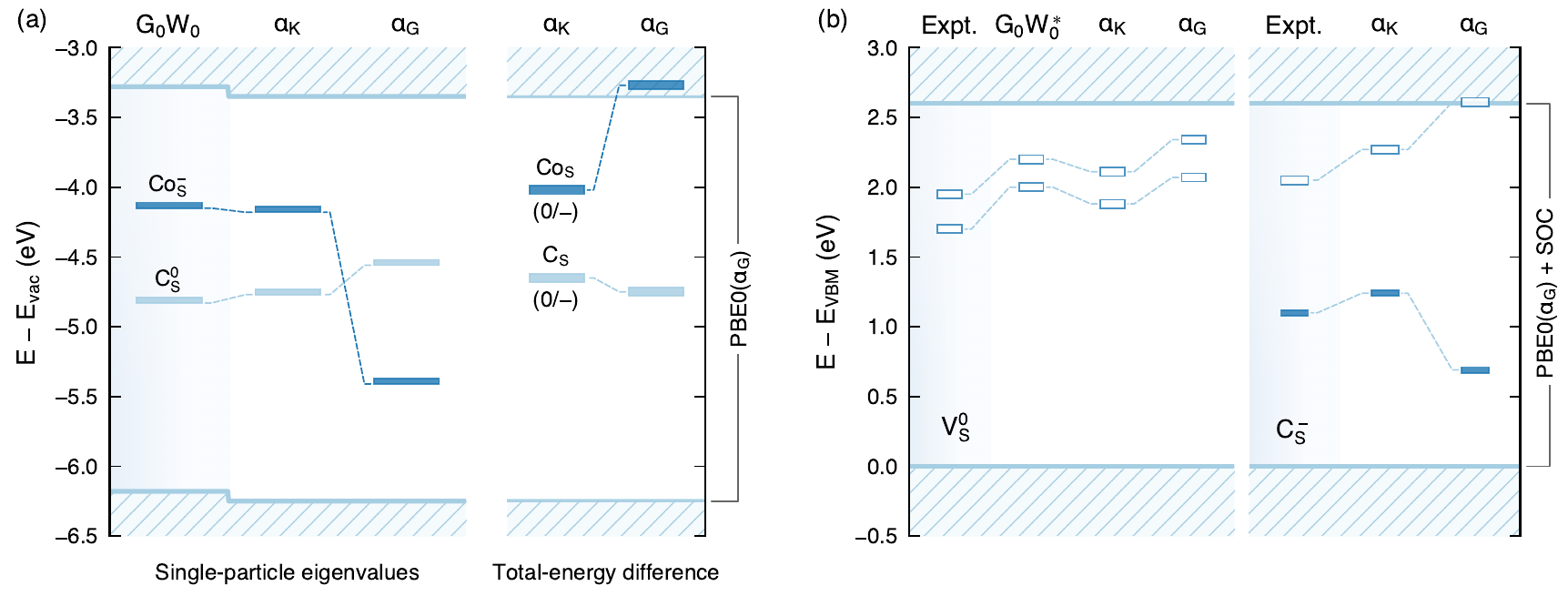}
\caption{\label{fig:g0w0_expt}(a) Defect levels of Co$_\text{S}^{-1}$ and C$_\text{S}^0$ obtained with
single-particle eigenvalues and total-energy difference with respect to the vacuum level.
For each computational scheme, the PBE0($\alpha$) defect levels are calculated with $\alpha_\text{K}$ and $\alpha_\text{G}$
and are benchmarked against the $G_0W_0$ reference.
To exclude the effect of structural relaxation, we use the PBE equilibrium structures throughout.
The PBE0 band-edge positions are determined by PBE0($\alpha_\text{G}$).
(b) Calculated single-particle defect levels with PBE0($\alpha_\text{K}$) and PBE0($\alpha_\text{G}$) compared to STS measurements for
$V_\text{S}^0$ and C$_\text{S}^-$.
The structures are relaxed with the corresponding $\alpha$ values for the PBE0 calculations.
The defect levels are referred to the VBM and take into account the SOC.
The band gap takes the value as obtained with PBE0($\alpha_G$)+SOC throughout.
The $G_0W_0$ results (denoted by $G_0W_0^\ast$) are taken from Ref.~\onlinecite{Schuler2019}.}
\end{figure*}

Figure~\ref{fig:be_ks} depicts the evolution of band edges and single-particle defect eigenvalues 
with respect to the mixing parameter $\alpha$.
The defect eigenvalues are obtained by linear interpolations or extrapolations
based on the eigenvalues calculated at four $\alpha$ values (0, 0.05, 0.10, and 0.22).
At a given $\alpha$, the defect structure upon which the Koopmans' condition is assessed 
corresponds to the equilibrium structure of the charged state.
At this stage, we do not take into account SOC for geometric relaxations 
and hence the determination of $\alpha_\text{K}$ due to its computational complexity.
For all defects considered, the values of $\alpha_\text{K}$ fulfilling the Koopmans' condition
are found in a narrow range between 0.06 and 0.11, significantly lower than the band-gap enforced 
mixing parameter $\alpha_\text{G}$ of 0.22. 
We note that $\alpha_\text{K}$ values are largely unaffected by SOC
as the Koopmans' condition is still fulfilled within 0.1 eV when SOC is included (see Table~S1 of the SM).
This is our first key result. 
Indeed, PBE($\alpha_\text{G}$) exhibits strong concavity as the ionization energy $-\varepsilon^\text{HO}(q)$ 
is consistently larger than the electron affinity $-\varepsilon^\text{LU}(q+1)$ for any given defect~\cite{Bruneval2009}.
This is clearly at odds with the existence of a unique mixing parameter for bulk materials
which is able to describe the band gap reasonably well without compromising the Koopmans' condition 
for localized defects~\cite{Miceli2018,Bischoff2019,Yang2022,Deak2019}. 

Among all defects considered, the effect of mixing parameter is particularly notable for Co$_\text{S}$.
The defect levels associated with Co$_\text{S}$ are predominately characterized by the strongly localized Co-$3d$ states.
Specifically, the (0/$-$) and ($+$/0) transitions involve the $d_{x^2-y^2}$ and the $d_{z^2}$ orbitals of the Co atom, respectively.
Taking the neutral Co$_\text{S}^0$ for example,
we notice that the occupied $d_{z^2}$ and unoccupied $d_{x^2-y^2}$ defect levels
move sharply towards the host band edges as $\alpha$ increases and eventually merge into the 
valence and conduction band when $\alpha=\alpha_\text{G}$ (cf.\ Fig.~\ref{fig:be_ks}).
Consequently, PBE0($\alpha_\text{G}$) would predict that no localized defect levels exist within the band gap for Co$_\text{S}$, 
qualitatively at variance with the deep levels obtained with PBE0($\alpha_\text{K}$).

To settle the conflicting results of Co$_\text{S}$ due to the choice of $\alpha$,
we carry out one-shot $G_0W_0$ calculations for the spin-unpolarized Co$_\text{S}^-$ defect 
(see the SM for computational details).
{We find excellent agreement between the PBE0($\alpha_\text{K}$) and $G_0W_0$
defect levels (mainly of $d_{x^2-y^2}$ and $d_{xy}$ characters of Co) [cf.\ Fig.~\ref{fig:g0w0_expt}(a)]. 
Both methods place the single-particle defect level $-4.1$~eV below
the vacuum level, or equivalently about 2.1~eV above the VBM.}
By contrast, PBE0($\alpha_\text{G}$) overestimates the defect ionization energy by nearly 1.3~eV,
consistent with the strong concavity noted previously in Fig.~\ref{fig:be_ks}.
The high accuracy of PBE0($\alpha_\text{K}$) is further demonstrated for C$_\text{S}^0$ [cf.\ Fig.~\ref{fig:g0w0_expt}(a)],
while PBE0($\alpha_\text{G}$) still underperforms, albeit to a lesser extent than for Co$_\text{S}^-$.

An arguably more common approach to the determination of defect levels is through
the total-energy difference between different defect charge states~\cite{Freysoldt2014}.
When referred to the VBM of the pristine host, 
the (0/$-$) charge transition level of defect $D$ can be obtained as 
$E_\text{tot}(D^-)-E_\text{tot}(D^0)-E_\text{VBM}$.
Analogous to the single-particle level, the total energy of charged defects are ill-defined for periodic systems.
Here we apply the finite-size correction scheme of Komsa \textit{et al.}~\cite{Komsa2013,Komsa2014,*Komsa2018} 
as implemented in \textsc{slabcc}~\cite{FarzalipourTabriz2019} to the total energy of charged defect.
The computed (vertical) (0/$-$) charge transition levels are shown in Fig.~\ref{fig:g0w0_expt}(a) for Co$_\text{S}$ and C$_\text{S}$.
In general, defect levels obtained with the total-energy difference scheme are less sensitive to the amount of Fock exchange
when aligned to a common reference (e.g., average electrostatic potential or vacuum)~\cite{Alkauskas2008,Komsa2010,Freysoldt2016,Lyons2017}.
While this is the case for the (0/$-$) level of C$_\text{S}$,
for Co$_\text{S}$ the agreement with the $G_0W_0$ QP energy found with PBE0($\alpha_\text{K}$) 
is substantially worse if $\alpha_\text{G}$ is used instead.

Our results so far suggest that fixing the $\alpha$ value, 
which is typically chosen based on the host band gap for studying defects in bulk materials, 
can be problematic and even qualitatively change the defect physics in 2D materials.
The discernible discrepancy between $\alpha_\text{G}$ and $\alpha_\text{K}$ suggests that defect levels need to be 
treated individually from band edges, the alignment of which can be easily achieved through the common vacuum level.
To illustrate the validity the proposed scheme, we refer to to the scanning tunneling spectroscopy (STS) 
measurements of the neutral $V_\text{S}^0$~\cite{Schuler2019} and the negatively charged C$_\text{S}^-$~\cite{Cochrane2021} 
in Fig.~\ref{fig:g0w0_expt}(b).
To facilitate the comparison, we include SOC in the hybrid-functional calculations and focus on the 
single-particle eigenvalues with the equilibrium defect structure at the charge state pertinent to experiment. 
The STS spectra revealed two prominent peaks at 1.8~eV and 2.0~eV above the VBM for $V_\text{S}^0$~\cite{Schuler2019}.
The peaks correspond to the unoccupied $d_{xy}$ and $d_{x^2-y^2}$ orbitals of the W atoms near the vacancy,
whose degeneracy is lifted by spin-orbit splitting~\cite{Li2016}.
The positions of the two defect levels are well reproduced 
by the single-particle defect energies of our PBE0($\alpha_\text{K}$) calculations
in reference to the band edges obtained with PBE0($\alpha_\text{G}$),
and the accuracy is comparable to that of the $G_0W_0$ calculation of Ref.~\cite{Schuler2019}.
The C$_\text{S}^-$ is also characterized by two defect levels deep within the band gap~\cite{Cochrane2021}.
The two levels are mainly of the C-$p_z$ character, with one 
occupied at 1.1~eV and the other unoccupied at 2.0~eV above the VBM.
The importance of enforcing the Koopmans' condition is again evidenced for C$_\text{S}^-$ as the splitting 
of the two defect levels would be too large if the mixing parameter is fixed at $\alpha_\text{G}$.

\begin{figure}
\includegraphics{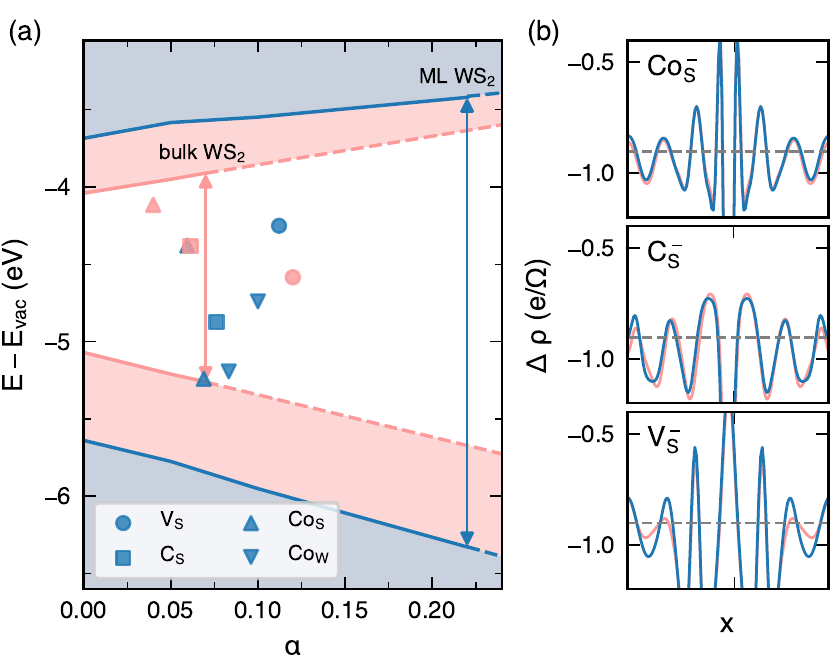}
\caption{\label{fig:be_scr}(a) Single-particle defect levels at the Koopmans-fulfilled $\alpha_\text{K}$
in the monolayer (blue) and the bulk (red) \ce{WS2}. 
The band-edge positions of the monolayer and the bulk \ce{WS2} are shown accordingly by the same color scheme.
(b) Defect charge density $\rho$($D^q$)$-\rho$($D^0$) of Co$^-_\text{S}$, C$^-_\text{S}$, and $V^-_\text{S}$ 
as obtained with PBE along the in-plane direction 
in the monolayer (blue) and the bulk (red) \ce{WS2}.
The horizontal line refers to the delocalized screening charge density $(1-1/\bar{\epsilon})q/\Omega$ given a localized defect
at charge state $q$ where $\bar{\epsilon}$ stands for the averaged dielectric constant of the bulk \ce{WS2}.
The defects are situated near the center.}
\end{figure}

To rationalize the departure from the existence of a unique $\alpha$ value as for bulk materials,
we turn to the bulk \ce{WS2} and examine the defect levels therein.
The bulk 2\textit{H}-\ce{WS2} comprises repeated layers of ML \ce{WS2}, 
and has an indirect ($\Gamma$--$K$) band gap of 1.3~eV~\cite{Kam1982,Gusakova2017}.
The markedly smaller band gap leads to a strongly reduced mixing parameter $\alpha_\text{G}^\text{bulk}$ of 0.07.
Defects are introduced to an orthorhombic supercell containing 72 atoms, which is based on
the 36-atom supercell of the ML \ce{WS2} comprising two alternating MLs stacked along the out-of-plane direction.
Figure~\ref{fig:be_scr}(a) shows the single-particle defect levels at the respective values of 
$\alpha_\text{K}$ for defects in both bulk and ML \ce{WS2}. 
Some defects such as Co$_\text{W}$ no longer exhibit deep levels in the bulk \ce{WS2}.
Interestingly, irrespective of the dimensionality of the host material, 
the various $\alpha_\text{K}$ values fall closely 
in the vicinity of $\alpha_\text{G}^\text{bulk}$.
The consistency in the $\alpha_\text{K}$ values implies that the localization of defect in the ML \ce{WS2} is comparable
to that of the bulk.
This is clearly demonstrated in Fig.~\ref{fig:be_scr}(b) where the defect charge density 
$\rho(D^q)-\rho(D^0)$ is plotted for the ML and the bulk host.
Particularly for Co$_\text{S}^-$ and C$_\text{S}^-$, the charge density along the in-plane direction 
is nearly intact when going from the bulk to the ML \ce{WS2}.
The charge density analysis additionally provides an avenue to understand the screening effect.
In the presence of a strictly localized defect, the average charge density approaches the delocalized 
screening charge density $(1-1/\epsilon)q/\Omega$ far away from the defect~\cite{Komsa2012},
where $\epsilon$ is the dielectric constant and $\Omega$ the volume of the supercell.
Using the (geometrically) averaged dielectric constant ($\bar{\epsilon}=10.2$) of the bulk \ce{WS2},
we find that the delocalized screening charge is recovered
for Co$_\text{S}^-$ and C$_\text{S}^-$ [cf.\ Fig.~\ref{fig:be_scr}(b)].
While this is expected for localized defects in the bulk \ce{WS2},
the agreement found in the ML \ce{WS2} is a compelling evidence that the localized defects therein 
are still subject to the bulklike screening and can be adequately described 
by $\alpha_\text{G}^\text{bulk}$.
The deviation from the ideal screening charge is more visible for \textit{V}$_\text{S}$,
in accord with its more delocalized nature and
hence the larger deviation of $\alpha_\text{K}$ from $\alpha_\text{G}^\text{bulk}$ compared to the substitutional defects.

{In contrast to $\alpha_\text{K}$, the large $\alpha_\text{G}$ for the ML \ce{WS2} is 
inherent to the strong opening of band gaps for 2D materials.}
We note in passing that the long-range part of Fock exchange is critical in opening the band gap of 2D materials
given the asymptotic $1/r$ decay in vacuum.
In fact, the short-ranged Heyd–Scuseria–Ernzerhof (HSE) functional~\cite{Heyd2003} struggles to 
open up the band gap of the ML \ce{WS2} unless an excessively large mixing parameter is used 
(0.45 with the PBE structure or 0.55 if the structure is relaxed self-consistently).
However, such a large mixing parameter results in a highly distorted band structure (e.g., valence bandwidth)
and, more importantly, shifts the whole band edges by $-0.3$~eV compared to $G_0W_0$ and PBE0($\alpha_\text{G}$) 
(see the SM).
Nonetheless, for localized defects, the HSE defect levels are reasonably aligned to the vacuum level 
as the short-range bulklike screening dominates. 

In addition to the TMDs, we find that the distinct two-set values ($\alpha_\text{K}$ and $\alpha_\text{G}$) 
also apply to the ML hexagonal boron nitride (\textit{h}-BN) involving only 
simple $sp$ elements (see the SM).
While systems characterized by localized $d$ electrons are less amenable to treatment 
for fulfilling the Koopmans' condition~\cite{Ivady2013,Yang2022},
we emphasize that the non-uniqueness of $\alpha$ is a general attribute for hybrid-functional 
defect calculations in 2D materials.

{Before closing, we note that our study focuses on defect energy levels.
It remains an open question how the nonuniqueness of $\alpha$ affects other defect properties,
such as the defect formation energies, the optical transitions, and the phonons.
Our findings motivate further investigations of the possible implications for general defect physics in 2D materials.}

In conclusion, we have shown that using a fixed amount of Fock exchange in hybrid functionals 
could lead to qualitatively incorrect description of defects in 2D materials, 
in contrast with the established practice common to defect calculations in bulk materials.
The absence of a unique mixing parameter stems from the reduced screening for the delocalized 
band-edge states whereas the screening is bulklike for localized defects.
The distinct screening behaviors hence require the band edges and the defect states to be treated separately
through their own optimal mixing parameters. 
We expect these effects to be present in all 2D materials and they should be carefully taken into account to 
ensure accurate hybrid-functional defect computations.

\begin{acknowledgments}
This work was supported by the U.S. Department of Energy, Office of Science, Basic Energy Sciences 
in Quantum Information Science under Award Number DE-SC0022289. 
This research used resources of the National Energy Research Scientific Computing Center, 
a DOE Office of Science User Facility supported by the Office of Science of the U.S.\ Department of Energy 
under Contract No.\ DE-AC02-05CH11231 using NERSC award BES-ERCAP0020966. 
Computational resources were also provided by supercomputing facilities of UCLouvain (CISM) and
Consortium des Equipements de Calcul Intensif en F\'ed\'eration Wallonie-Bruxelles (CECI).
The authors acknowledge Y.\ Ping, M.\ C.\ da Silva, and P.\ De\'ak for fruitful discussions.
\end{acknowledgments}

\bibliography{main}
\end{document}


\title{\textsc{Supplemental Material}\\
Non-unique fraction of Fock exchange for defects in two-dimensional materials}
\author{Wei Chen}
\affiliation{Institute of Condensed Matter and Nanoscicence (IMCN), Universit\'{e} catholique de Louvain, Louvain-la-Neuve 1348, Belgium}
\author{Sin\'ead M. Griffin}
\affiliation{Molecular Foundry Division, Lawrence Berkeley National Laboratory, Berkeley, California 94720, USA}
\affiliation{Materials Sciences Division, Lawrence Berkeley National Laboratory, Berkeley, California 94720, USA}
\author{Gian-Marco Rignanese}
\affiliation{Institute of Condensed Matter and Nanoscicence (IMCN), Universit\'{e} catholique de Louvain, Louvain-la-Neuve 1348, Belgium}
\author{Geoffroy Hautier}
\affiliation{Institute of Condensed Matter and Nanoscicence (IMCN), Universit\'{e} catholique de Louvain, Louvain-la-Neuve 1348, Belgium}
\affiliation{Thayer School of Engineering, Dartmouth College, Hanover, New Hampshire 03755, USA}
\date{\today}

\maketitle

\section{\label{sec:comp}Computational details}
We employ two orthorhombic supercells to represent the monolayer (ML) \ce{WS2}. 
The majority of hybrid functionals are carried out with the 90-atom orthorhombic supercell by transforming the 
in-plane lattice vectors of the primitive cell through the matrix 
$\big(\begin{smallmatrix}
5 & 5 \\ -3 & 3 
\end{smallmatrix}\big)$.
The out-of-plane lattice parameter $L_z=24$ \AA.
A smaller 36-atom orthorhombic supercell is obtained via the transformation matrix 
$\big(\begin{smallmatrix}
3 & 3 \\ -2 & 2 
\end{smallmatrix}\big)$ with $L_z=16$ \AA.
The smaller supercell is used in $G_0W_0$ calculations and for analyzing the finite-size effect.
The experimental in-plane lattice constant of the bulk 2\textit{H}-\ce{WS2} ($a=3.15$ \AA~\cite{Schutte1987}) is used 
without further relaxation for the ML \ce{WS2}.

The kinetic energy cutoff is 340~eV for the projector-augmented wave (PAW) calculations.
For the carbon substitutional defect (C$_\text{S}$), the cutoff is increased to 400~eV.
We use a $k$-point mesh density equivalent to $12\times12\times1$ for the primitive cell of ML \ce{WS2} 
(e.g., $4\times4\times1$ for the 36-atom supercell and $3\times3\times1$ for the 90-atom supercell).
The evaluation of Fock exchange is sped up with the adaptively compressed exchange operator formulation~\cite{Lin2016}
with an energy cutoff of 270~eV (320~eV for C$_\text{S}$).
The atomic coordinates are relaxed until the residual forces are below 0.02~eV/\AA.

The mean-field starting point for the one-shot $G_0W_0$ calculations are carried out 
using the optimized norm-conserving pseudopotentials~\cite{Hamann2013} 
made available via the \textsc{PseudoDojo} project~\cite{vanSetten2018}.
We use the $4\times4\times1$ $\Gamma$-centered $k$-point mesh for the 36-atom orthorhombic supercell.
The dielectric matrix is evaluated with up to 5000 bands and a planewave energy cutoff of 10~Ry.
Frequency dependence of dielectric function is taken into account by the Godby-Needs plasmon-pole model~\cite{Godby1989}.
The truncated Coulomb interaction is applied along the out-of-plane direction~\cite{IsmailBeigi2006}.
The slow convergence with respect to $q$ points is addressed by the subsampling technique~\cite{Jornada2017},
which includes 3 additional $q$ points for $q\rightarrow0$.
Spin-orbit coupling is not included in $G_0W_0$ calculations.

{We find the QP energies can be reasonably described with a smaller
orthorhombic supercell of 36 atoms owing to the strong localization of the defect.
For results depicted in Fig.~2(a) of the main text,
we use the PBE equilibrium configuration of the neutral defect
and refrain from further relaxations for the negatively charged defect
in order to exclude the structural effect.}

\section{\label{sec:correction}Finite-size corrections}
\begin{figure}
\includegraphics{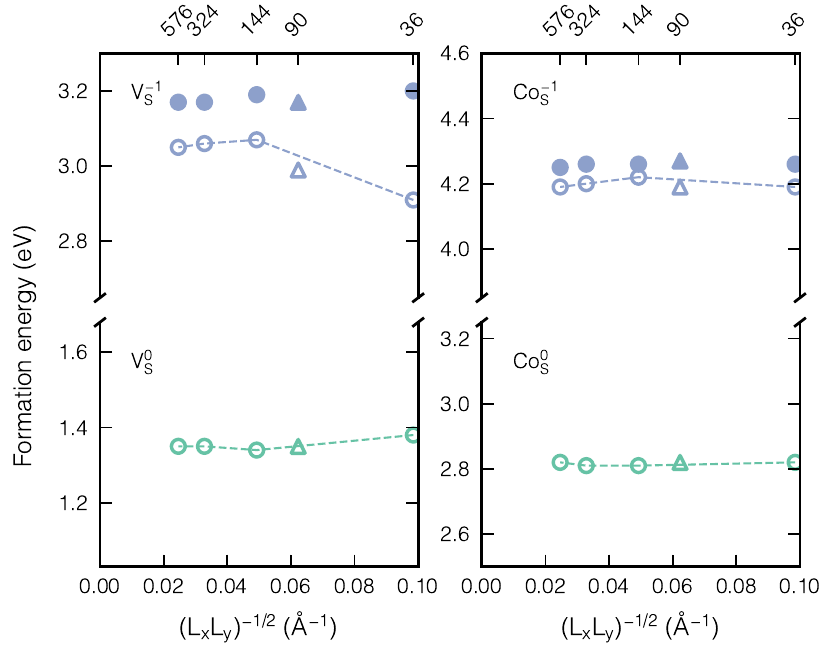}
\caption{\label{fig:corr_energy}Uncorrected and corrected formation energies for the 
\textit{V}$_\text{S}$ and Co$_\text{S}$ defects calculated with DFT-PBE. 
For the series of supercells with 36, 144, 324, and 576 atoms (denoted by circles) 
the supercell length $L_z=1.6(L_xL_y)^{1/2}$. 
For the 90-atom supercell (denoted by triangles) $L_z=1.5(L_xL_y)^{1/2}$.
For the charged defects, the uncorrected (corrected) formation energies are shown by the empty (filled) circles.} 
\end{figure}

The total energies of charged point defects are corrected by the
\textit{a posteriori} correction scheme~\cite{Komsa2013,Komsa2014,FarzalipourTabriz2019}.
Since ionic contributions to the dielectric screening is negligible for \ce{WS2},
we use the high-frequency dielectric constant throughout.
Figure~\ref{fig:corr_energy} shows the formation energies of 
\textit{V}$_\text{S}$ and Co$_\text{S}$ in the ML \ce{WS2}.
Co$_\text{S}$ exhibits strongly localized defect wavefunctions whereas \textit{V}$_\text{S}$ is less localized.
Overall the corrected formation energies for the charged defects are converged already with the 36-atom supercell regardless of defect localization.

\begin{figure}
\includegraphics{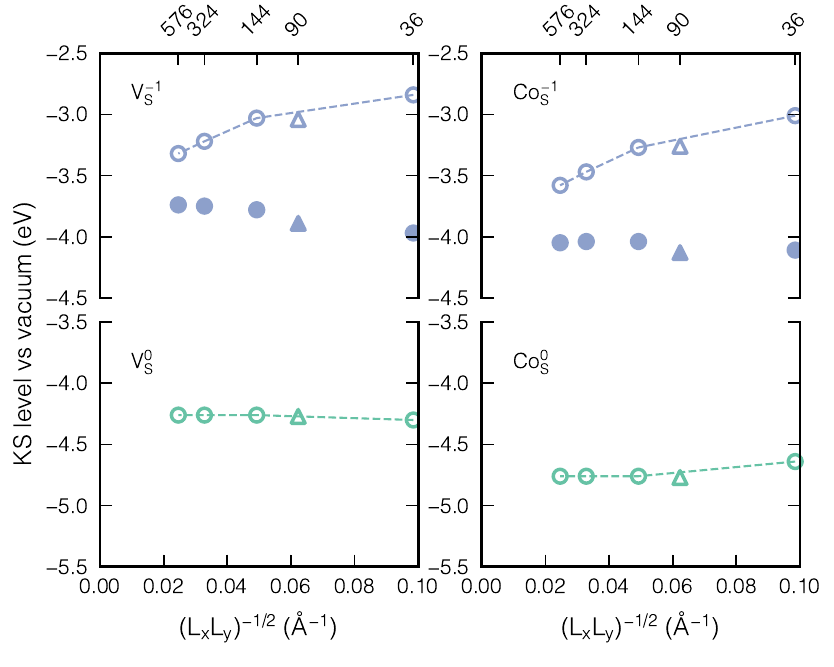}
\caption{\label{fig:corr_eigenvalue}Uncorrected and corrected KS single-particle eigenvalues
with respect to the vacuum level calculated with DFT-PBE.}
\end{figure}

For charged defects in the ML \ce{WS2}, the electrostatic potential and eigenvalues are corrected concurrently using 
the scheme of Chagas da Silva \textit{et al.}~\cite{ChagasdaSilva2021}.
Figure~\ref{fig:corr_eigenvalue} shows the Kohn-Sham (KS) single-particle eigenvalues associated with 
\textit{V}$_\text{S}$ and Co$_\text{S}$.
The electrostatic potential furthest away from the defect is taken as the vacuum reference level.
While the single-particle eigenvalues are well defined for the neutral defects,
they are subject to strong finite size effect as the supercell size varies.
The potential correction scheme works well for the negatively charged Co$^{-}_\text{S}$ 
and the corrected eigenvalue is already converged within 0.1~eV with the smallest 36-atom supercell.
For \textit{V}$^-_\text{S}$ the slower convergence of the corrected eigenvalues can be attributed 
to the more delocalized nature of the defect wavefunctions.
Nonetheless, the eigenvalue of \textit{V}$^-_\text{S}$ is still reasonably converged within 0.2~eV with the 36-atom supercell.

For defects in the bulk \ce{WS2}, we apply the eigenvalue correction $-\frac{2}{q}E_\text{corr}$ 
where $E_\text{corr}$ is the total-energy correction~\cite{Chen2013}.

\section{Effect of structural relaxations}
In Fig.~2a of the main text, the single-particle defect levels are obtained using the neutral charge defect structures relaxed at 
the PBE level.
Here we investigate the effect of structural relaxations on the single-particle defect levels.
Taking Co$^0_\text{S}$ for example, we show in Table~\ref{tab:relax} that relaxing the structure using 
PBE0($\alpha_\text{K}=0.06$) increases the Co--W bond lengths 
by 0.02~\AA under the $C_S$ symmetry.
The slighlt modification in the defect structure leads to a marginal shift of the defect energy level by 0.04 eV.

We note that the comparison as done in Fig.~2a is nonethless unaffected by the choice of the reference structure.

\begin{table}
\caption{\label{tab:relax}Difference in the nearest Co--W bond lengths and in the single-particle defect level 
due to the choice of the reference Co$^0_\text{S}$ structure. 
The two structures, referred to as R($\alpha$), correspond
to the equilibrium structure obtained with PBE0($\alpha$). 
The single-particle defect level is evaluated with PBE0($\alpha_\text{K}=0.06$) and is referred to the vacuum level.}
\begin{ruledtabular}
\begin{tabular}{lccc}
R($\alpha$)               & $d_{\text{Co}-\text{W}_1}$ (\AA) & $d_{\text{Co}-\text{W}_{2,3}}$ (\AA) & $\epsilon^\text{HO}(-)$ (eV)\\
\hline
R(0)                      & 2.48                 & 2.52                     & -4.16 \\
R($\alpha_\text{K}=0.06$) & 2.50                 & 2.54                     & -4.12 \\
\end{tabular}
\end{ruledtabular}
\end{table}

\section{Role of long-range Fock exchange for 2D systems}
\begin{figure}
\includegraphics{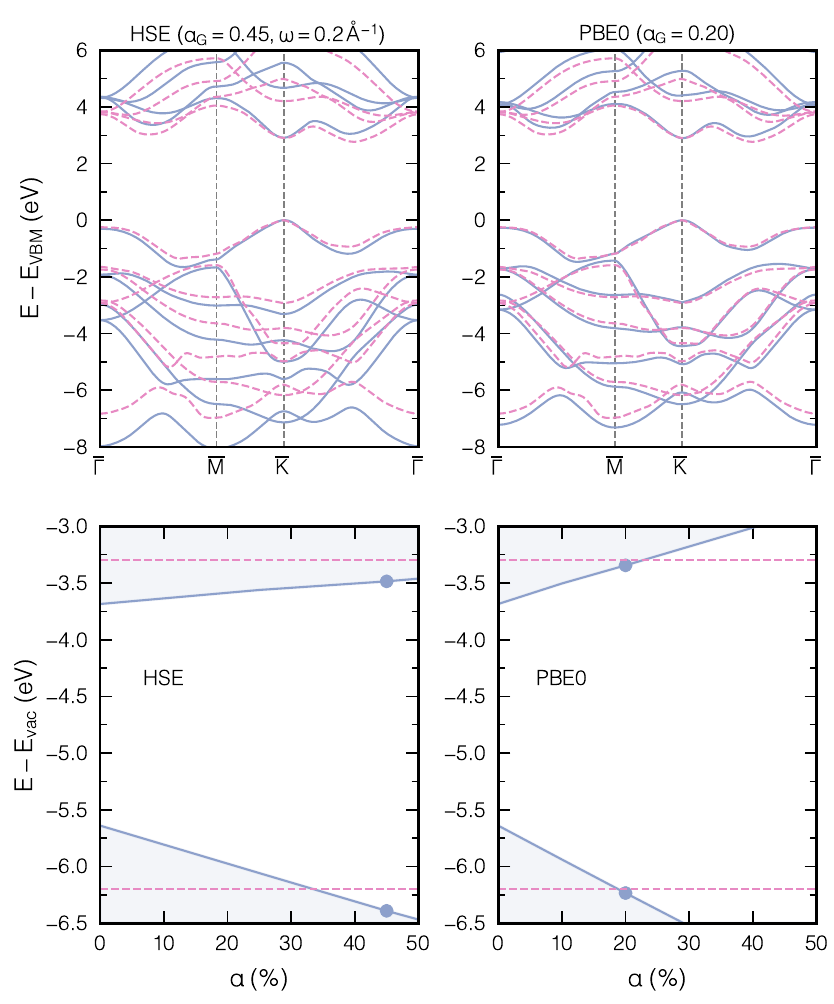}
\caption{\label{fig:bs}Band structure of ML WS$_2$ (top panel) and absolute band-edge positions (bottom) 
obtained with the short-range hybrid functional (HSE) and the global hybrid functional retaining the long-range part of the Fock exchange (PBE0). 
In the band-structure plot both functionals use a mixing parameter $\alpha$ leading to the $G_0W_0$ band gap.  
The $G_0W_0$ band structure is shown in dashed lines. 
The energy is referred to the top of the VBM at $K$.
The band-edge positions are shown as a function of the mixing parameter $\alpha$.
The $\alpha$ value at which the $G_0W_0$ band gap is reproduced is highlighted.
The $G_0W_0$ band-edge positions are indicated by the dashed lines.
}
\end{figure}

The range-separated HSE hybrid functional only retains the Fock exchange in the short range.
This is practical for 3D semiconductors as it accelerates the computation, 
but for 2D systems the short-range hybrid functional is insufficient to account for the band-gap opening at the reduced dimension.
While it is in principle possible to recover the band gap with an exceedingly large $\alpha$ in HSE,
the band structure is visibly distorted.
This is manifested, for instance, by the too large VBM bandwidth and the overestimated 
direct band gap at $\Gamma$ (cf.\ Fig.~\ref{fig:bs}).
On the other hand, the band structure of PBE0($\alpha_\text{G}$) is in much better agreement with the $G_0W_0$ one.
In addition, the large $\alpha$ places the HSE band edges about 0.3~eV too deep compared to the $G_0W_0$,
PBE0($\alpha_\text{G}$) again achieves an excellent agreement in terms of the absolution band-edge positions.

\begin{figure}
\includegraphics{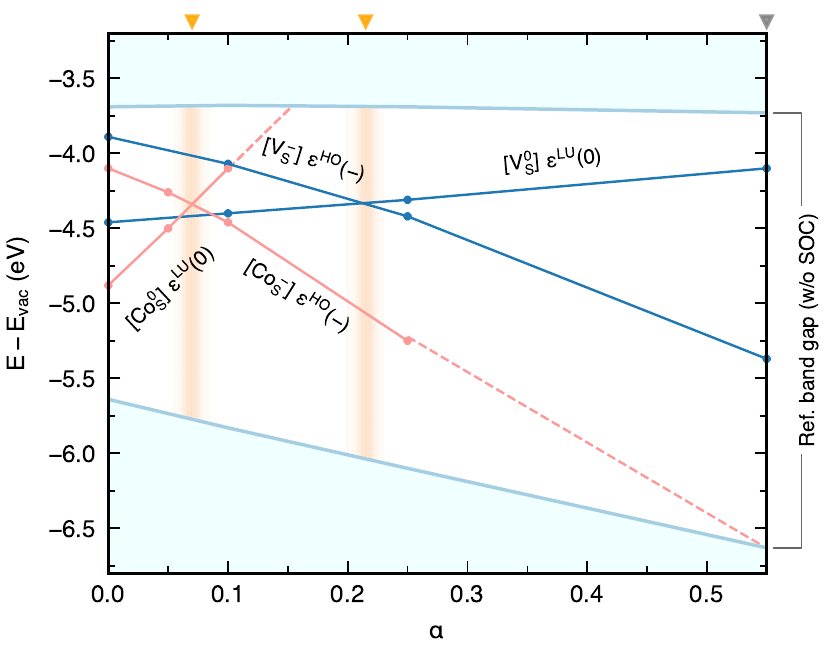}
\caption{\label{fig:be_ks_hse}{Single-particle defect levels as a function of $\alpha$ with the range-separated 
HSE functional. The inverse screening length is set to 0.2 \AA$^{-1}$. For legends refer to Fig.~1 of the main text.}}
\end{figure}

\begin{figure}
\includegraphics{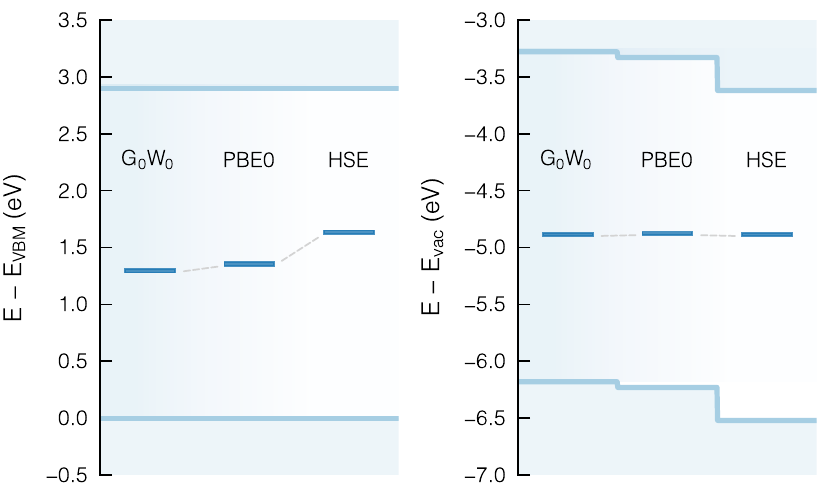}
\caption{\label{fig:cos_g0w0}Single-particle defect level of Co$^-_\text{S}$ calculated with two families of hybrid functionals (PBE0
and HSE) and compared to $G_0W_0$ reference. The defect levels are referred to the VBM (left) and to the vacuum level (right).
For the hybrid functionals, the band edges are obtained with the optimal mixing parameter ($\alpha_\text{G}$) 
which reproduces the $G_0W_0$ band gap, 
whereas the defect levels are determined by the mixing parameter ($\alpha_\text{K}$) fulfilling Koopmans' condition.
Concretely, $\alpha_\text{G}$ is 0.22 (0.55) for PBE0 (HSE), and $\alpha_\text{K}$ is 0.07 (0.07) for PBE0 (HSE).
The inverse screening length is fixed at 0.2~\AA$^{-1}$ for HSE.}
\end{figure}

The too deep VBM has an implication for interpreting defect levels when referenced to the band edges.
As shown in Fig.~\ref{fig:cos_g0w0}, the HSE single-particle defect level of Co$^-_\text{S}$ is about 0.3~eV 
higher than those of PBE0 and $G_0W_0$ as a result of the misaligned VBM.
Nevertheless, HSE still leads to consistent descriptions of defect levels insofar as the vacuum level 
is used as the common reference.

Figure~\ref{fig:be_ks_hse} shows the single-particle defect levels as a function of $\alpha$ with the range-separated HSE functional.
We obtain $\alpha_\text{K}=0.21$ for V$_\text{S}^{0/-}$ and $\alpha_\text{K}=0.07$ for Co$_\text{S}^{0/-}$.
In accord with the PBE0 results, these values are considerably smaller than the $\alpha_\text{G}$ (0.55).
The sizable discrepancy between $\alpha_\text{K}$ and $\alpha_\text{G}$ is therefore universal for 2D materials 
regardless of range separation.

\section{Koopmans' condition and spin-orbit coupling}
{The values of $\alpha_\text{K}$ (i.e., the mixing parameter fulfilling the Koopmans' condition) 
have been determined without taking into account spin-orbit coupling (SOC) in the main text.
We show in Table~\ref{tab:soc} that Koopmans' condition is still largely fulfilled at these $\alpha_\text{K}$ values
if the SOC is taken into account.
In particular, the two single-particle eigenvalues agree within 0.1 eV for all the defects considered in this work.}

\begin{table}
\caption{\label{tab:soc}{Single-particle eigenvalues (in eV) of the pertinent defect states with respect to the vacuum level
including the SOC effect. The fraction of Fock exchange at which the eigenvalues are evaulated follows the $\alpha_\text{K}$ values
from the main text without taking into account SOC. 
For any given defect, the equilibrium structure of the charged defect does not incude the SOC effect 
and is used for the neutral one.}}
\begin{ruledtabular}
\begin{tabular}{ldddddd}
 &
\multicolumn{1}{c}{V$^{-1}_\text{S}$} &
\multicolumn{1}{c}{C$^{-1}_\text{S}$} &
\multicolumn{1}{c}{Co$^{-1}_\text{W}$} &
\multicolumn{1}{c}{Co$^{+1}_\text{W}$} &
\multicolumn{1}{c}{Co$^{-1}_\text{S}$} &
\multicolumn{1}{c}{Co$^{+1}_\text{S}$} \\[0.5ex]
\hline
\rule{0pt}{3ex}$\varepsilon(q)$ & -4.25 & -4.81 &  -4.69 & -5.23  & -4.17 & -5.23\\
$\varepsilon(0)$ & -4.18 & -4.88 &  -4.76 & -5.32  & -4.26 & -5.23\\
\end{tabular}
\end{ruledtabular}
\end{table}

\section{Koopmans' condition for monolayer boron nitride}
\begin{table}
\caption{\label{tab:hbn}Highest occupied eigenvalues of the negatively charged C$_\text{N}^{-}$ [$\varepsilon^\text{HO}(-)$]
and lowest unoccupied eigenvalues of the neutral C$_\text{N}^0$ [$\varepsilon^\text{LU}(0)$]
obtained with $\alpha_\text{G}^\text{ML}$ along with a vanishing $\alpha$ value and the standard $\alpha=0.25$.
The equilibrium structure of C$_\text{N}^{-}$ obtained with the PBE functional is used throughout.
The eigenvalues are referred to the vacuum level and are in eV.
The deviation from piecewise linearity is indicated by $\Delta$.
A positive (negative) $\Delta$ value corresponds to convexity (concavity).}
\begin{ruledtabular}
\begin{tabular}{lddd}
             & \multicolumn{1}{c}{$\alpha=0$} & 
               \multicolumn{1}{c}{$\alpha=0.25$} &
               \multicolumn{1}{c}{ $\alpha_\text{G}^\text{ML}=0.35$} \\[0.5ex]
\hline
\rule{0pt}{3ex}$\varepsilon^\text{HO}(-)$ & -3.50 & -4.34  &  -4.70 \\   
$\varepsilon^\text{LU}(0)$ & -5.11 & -4.42  &  -4.08 \\
$\Delta$                   &  1.61 &  0.08  &  -0.62 \\
\end{tabular}
\end{ruledtabular}
\end{table}

We extend the analysis to the hexagonal boron nitride (\textit{h}-BN), a simple $sp$ semiconductor.
We use the C$_\text{N}$ substitutional defect to probe to what extent the Koopmans' condition is fulfilled at 
given $\alpha$ values.
The band gap of the ML \textit{h}-BN, which corresponds to the indirect $K-\Gamma$ transition,
is 7.0~eV according to the $G_0W_0$ calculations of Ref.~\cite{Smart2018}.
To reproduce this $G_0W_0$ band gap, the mixing parameter $\alpha_\text{G}^\text{ML}$ of 0.35 is needed for
the PBE0 hybrid functional.
In comparison, the standard PBE0($\alpha=0.25$) describes the band gap of the bulk \textit{h}-BN reasonably well.

The defect calculations are carried out with a 72-atom orthorhombic supercell with $L_z=15$~\AA\ using 
an energy cutoff of 400~eV and a $\Gamma$-centered $4\times4\times4$ $\mathbf{k}$-point mesh.
Table~\ref{tab:hbn} shows the highest occupied and the lowest unoccupied eigenvalues pertinent to the
determination of Koopmans' condition at three representative $\alpha$ values, namely 0, 0.25, and $\alpha_\text{G}^\text{ML}$.
As expected, the Koopmans' condition is best satisfied at $\alpha=0.25$, the value closely reproducing the band gap 
of the bulk \textit{h}-BN.
The other two values lead to either strong convexity or strong concavity.
The Koopmans' $\alpha_\text{K}$ value is estimated to be 0.26, significantly lower than the $\alpha_\text{G}^\text{ML}$. 
This shows that the non-uniqueness of $\alpha$ is a general attribute applicable to a wide class of 2D materials.

\section{Data availability}
{The input and output files for the defect calculations presented in this work are available at 
\url{https://doi.org/10.5281/zenodo.6916999}.}

\bibliography{main}